\documentclass[11pt,twoside]{article}

\usepackage{asp2006_astroph}
\usepackage{epsf}
\usepackage{lscape}
\usepackage{graphicx}

\begin{document}
\title{Extreme Outer Galaxy: A Laboratory of Star Formation in an
 Early Epoch of Galaxy Formation} 

\author{Naoto Kobayashi\altaffilmark{1}, Chikako Yasui\altaffilmark{1},
Alan T. Tokunaga\altaffilmark{2}, and Masao Saito\altaffilmark{3}}


\altaffiltext{1}{IoA, University of Tokyo, 2-21-1
Osawa, Mitaka, Tokyo 181-0015, Japan}

\altaffiltext{2}{IfA, University of Hawaii, 2680
Woodlawn Drive, Honolulu, HI 96822, USA}

\altaffiltext{3}{ALMA Project, NAOJ, 2-21-1 Osawa, Mitaka, Tokyo
181-8588, Japan}



\begin{abstract} 

The extreme outer Galaxy (EOG) has a very different environment from
that in the solar neighborhood, with low metallicity (less than $-0.5$
dex), much lower gas density, and small or no perturbation from spiral
arms. The EOG is an excellent laboratory for the study of the star
formation processes that happened during the formation period of the
Galaxy. In particular, {\it the study of the EOG may shed light on the
origin and role of the thick disk, whose metallicity range matches well
with that of the EOG}. We show an example of a molecular cloud in the
EOG (Digel's Cloud 2), which is located at $R\mathrm{_g} \sim 20$ kpc
beyond the Outer arm. Based on our NIR and $^{12}$CO data as well as HI,
radio continuum, and IRAS data in the archives, we examined the detailed
star formation processes in this unique environment, especially the
supernova triggered star formation, which should have been the major
star formation mode during the halo and thick disk formation.

\end{abstract}


\section{Introduction \,\,\,$-$ Connection with Galaxy Formation $-$}

\begin{minipage}[l]{6.9cm}
   \begin{flushleft}
The extreme outer Galaxy (EOG) region, which we define as the region at
the Galactic radius $R\mathrm{_g} > 18$ kpc (Fig 1), has a very
different environment from the solar neighborhood since it has much
lower gas density, lower metallicity and small or no perturbation from
spiral arms. Such a region is not only of strong interest in itself, but
it also provides an opportunity to study astronomical processes, such as
star formation in a physical environment that is much different from
that in the solar neighborhood. While the detailed star formation
processes have been studied mostly for nearby star forming regions at
the distances less than 1 kpc, where
   \end{flushleft}
\end{minipage}
\hspace{3mm}
\begin{minipage}{6.5cm}
\vspace{1mm}
   \begin{center}
     \includegraphics[width=6.0cm]{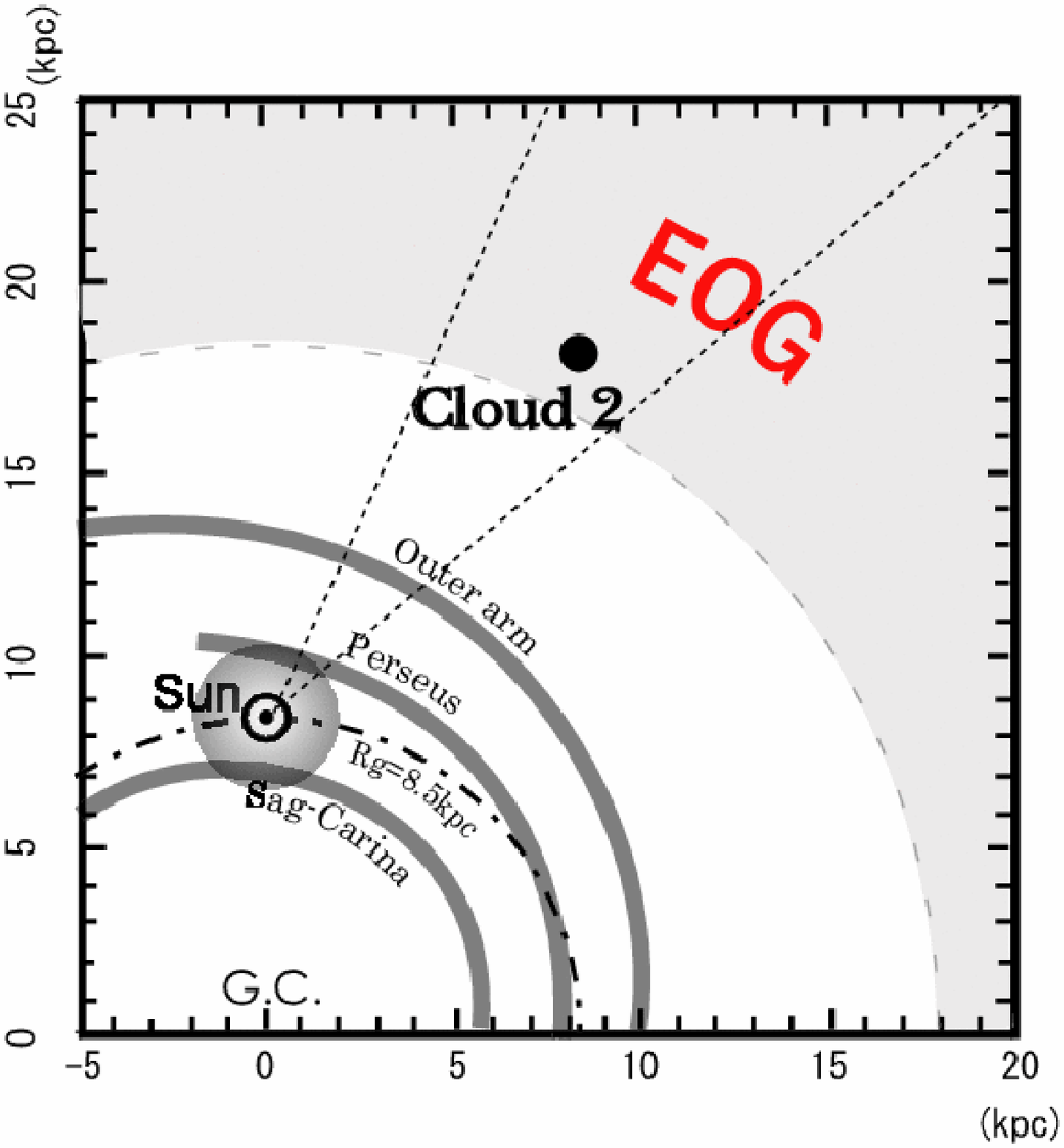}
   \end{center}
\vspace{-3mm}
 {\label{fig:bestIMF} {\bf Figure 1.} {%
EOG and the location of Cloud 2.
}}
\end{minipage}

\noindent 
the physical/chemical environment is relatively uniform, the EOG enables
us to study how the environment (density, temperature, pressure,
external radiation field, metallicity, etc) affects the basic star
formation processes and parameters such as IMF, star formation
efficiency, and disk formation efficiency.

The metallicity of outer galaxy disks as well as of Im galaxies,
low-metallicity blue compact dwarf galaxies, and damped Lyman-$\alpha$
systems is observed to be $[M/H] \sim -0.5$ to $-3$. This metallicity
range traces the critical phases of early Galaxy formation when the
major components, such as halo, thick disk, bulge, and thin disk, were
formed \citep{FB2002}. The typical metallicity of the outer galaxy
regions ($-1.5 < [M/H] < -0.5$) suggests that they represent the late
phase of the Halo formation and the early phase of the thick disk
formation.  The observational study of the outer Galaxy region may
reveal details of the star formation process during the formation of the
thick disk, which took place about 10 Gyrs ago, and may shed light on
the origin and role of this important galactic component.


\section{A Case Study \,\,\,$-$ Digel Cloud 2 $-$}

In order to study the star formation in the EOG, we are conducting a
near-infrared survey of the Digel clouds, which are one of the best
samples of molecular clouds in the EOG \citep{D1994}.
Cloud 2 is the most massive molecular cloud among the Digel clouds at
the probable galactic distance $R\mathrm{_g} \sim 19$ kpc (Fig 1).
After the initial discovery of young NIR objects in Cloud 2
\citep{KT2000}, we obtained deeper NIR images using the QUIRC NIR-imager
at the University of Hawaii 2.2 m telescope to find two young embedded
clusters in this cloud.
We also found a clear signature of sequential star formation in the
expansion direction of an adjacent supernova (SN) shell, GSH-138-01-94
\citep{SI2001}, suggesting that the star formation was triggered by this
huge SN shell with an estimated age of 4.3 Myr. From the projected
angular distance between the clusters and the shell, the age of the
clusters can be estimated assuming that the passage of the SN shell
started the cluster formation: it is less than 1.2 Myr but most likely
$\sim$ 0.5 Myr considering the location of the cluster in the line of
sight.
It is thought that consecutive SN-triggered star formation was the
major star formation process in the early epoch of galaxy formation
because of the lack of other major star formation triggers such as the
density waves in spiral galaxies \citep[e.g.,][]{T1999}. Further
detailed study of the Cloud 2 clusters with deep NIR imaging are
presented in \citet{Y2006,Y2008} (see also Yasui et al. in this volume).





\end{document}